\documentclass[preprint,12pt]{elsarticle}

\usepackage[OT4]{fontenc}

\usepackage[utf8]{inputenc} 

\usepackage[breaklinks=true]{hyperref}
\usepackage{algorithm}
\usepackage{graphicx}
\usepackage[noend]{algpseudocode}

\usepackage{amssymb}
\usepackage{array}
\usepackage{booktabs}
\usepackage[usenames,dvipsnames]{color}
\usepackage{enumitem}
\usepackage{caption}
\usepackage{adjustbox}
\usepackage{setspace}
\usepackage[section]{placeins}
\usepackage{gensymb}
\usepackage{mathtools}
\usepackage{amsthm}
\usepackage{lineno}
\usepackage{rotating}
\usepackage{caption}
\usepackage{subcaption}

\usepackage{tabularx}
\usepackage{multirow}

\algnewcommand{\LeftComment*}[2]{\Statex \hspace{#1}$\triangleright$ #2}
\renewcommand*\Call[2]{\textproc{#1}(#2)}

\journal{arXiv}

\newcommand{\algo}{RuleKit-CS}

\newcommand{\mincov}{\textit{minsupp-new}}
\newcommand{\MINCOVall}{\textit{MINSUPPs}}
\newcommand{\mincovall}{\textit{minsupp-all}}
\newcommand{\maxneg}{\textit{max-neg2pos}}
\newcommand{\maxpass}{\textit{max-passes}}
\newcommand{\qual}{\textit{quality}}

\newcommand{\penal}{\pi}
\newcommand{\reward}{\phi}
\newcommand{\cand}{\ \wedge\ }

\newcommand{\ii}[1]{\textit{#1}}

\newcommand{\AND}{\ \textbf{and}\ }
\newcommand{\OR}{\ \textbf{or}\ }
\newcommand{\IN}{\ \textbf{in}\ }

\newcommand{\e}{\phantom{0}}

\newcommand{\ts}{\textsuperscript}

\newcommand{\link}[1]{\url{#1}}

\begin{document}
	
	\begin{frontmatter}
		
		\title{Separate and conquer heuristic allows robust mining of contrast sets in classification, regression, and survival data}
		
		
		\author[polsl,emag]{Adam Gudy\'s\corref{mycorrespondingauthor}}
		\ead{adam.gudys@polsl.pl}
		
		\author[polsl,emag]{Marek Sikora\corref{mycorrespondingauthor}}
		\ead{marek.sikora@polsl.pl}
		
		\author[polsl,emag]{\L{}ukasz Wr\'obel}
		\ead{lukasz.wrobel@polsl.pl}

		\cortext[mycorrespondingauthor]{Corresponding author}
		\address[polsl]{Faculty of Automatic Control, Electronics and Computer Science, Silesian University of Technology, Akademicka 16, 44-100 Gliwice, Poland}
		
		\address[emag]{Institute of Innovative Technologies, EMAG, Leopolda 31, 40-189 Katowice, Poland}

\begin{abstract}
Identifying differences between groups is one of the most important knowledge discovery problems. The procedure, also known as contrast sets mining, is applied in a wide range of areas like medicine, industry, or economics. 

In the paper we present \algo{}, an algorithm for contrast set mining based on separate and conquer  -- a well established heuristic for decision rule induction. Multiple passes accompanied with an attribute penalization scheme provide contrast sets describing same examples with different attributes, distinguishing presented approach from the standard separate and conquer. The algorithm was also generalized for regression and survival data allowing identification of contrast sets whose label attribute/survival prognosis is consistent with the label/prognosis for the predefined contrast groups. This feature, not provided by the existing approaches, further extends the usability of \algo{}.   

Experiments on over 130 data sets from various areas and detailed analysis of selected cases confirmed \algo{} to be a useful tool for discovering differences between defined groups. The algorithm was implemented as a part of the RuleKit suite available at GitHub under GNU AGPL~3 licence (\link{https://github.com/adaa-polsl/RuleKit}).
\end{abstract}


\end{frontmatter}


\section{Introduction}
In the knowledge discovery in tabular data, rules are the most intuitive, thus the most popular knowledge representation. As Novak et al.~\citep{novak2009supervised} noticed, a lot of knowledge discovery tasks can be considered as rule induction problems. These are, for instance, association rule learning~\citep{PiatetskyShapiro1991, agrawal1994}, subgroup discovery~\citep{klosgen1996,lavrac2004}, contrast set mining~\citep{bay2001detecting}, or identifying emerging patterns~\citep{dong1999efficient}. While descriptive capabilities of rules are indisputable, they can be also used for predictive purposes, i.e., for building classification systems~\citep{michalski1973discovering,clark1989,cohen1995fast}. Initially, these two directions were independently investigated by data mining (descriptive) and machine learning (predictive) communities~\citep{novak2009supervised}. The aforementioned perspectives are, however, tightly related. In fact, any rule induction algorithm can be oriented towards one or both of these purposes. The differences lie in the applied strategy of exploring the search space, the methods of assessing the rules, and their post-processing.

For knowledge discovery purposes, the induction often tries to find all rules fulfilling assumed quality constraints, e.g., precision (confidence) or support (coverage). This can be followed by the filtering based on the rule interestingness~\citep{geng2006interestingness}. When classification is the main aim, the induction is oriented towards highest predictive power. Ensemble of rules~\citep{gu2018massively} are particularly effective at this field. The possibility to interpret resulting models is often illusive, though. Among many rule learning approaches, separate and conquer~\citep{furnkranz1999} (also known as sequential covering) is a reasonable compromise allowing induction of a moderate number of rules with good predictive power. Importantly, the procedure can be straightforwardly adjusted towards interpretability or classification abilities of the model by using different rule quality measures~\citep{janssen2010quest,wrobel2016rule}.    

\begin{figure}
	\centering
	\includegraphics[width=1.0\textwidth]{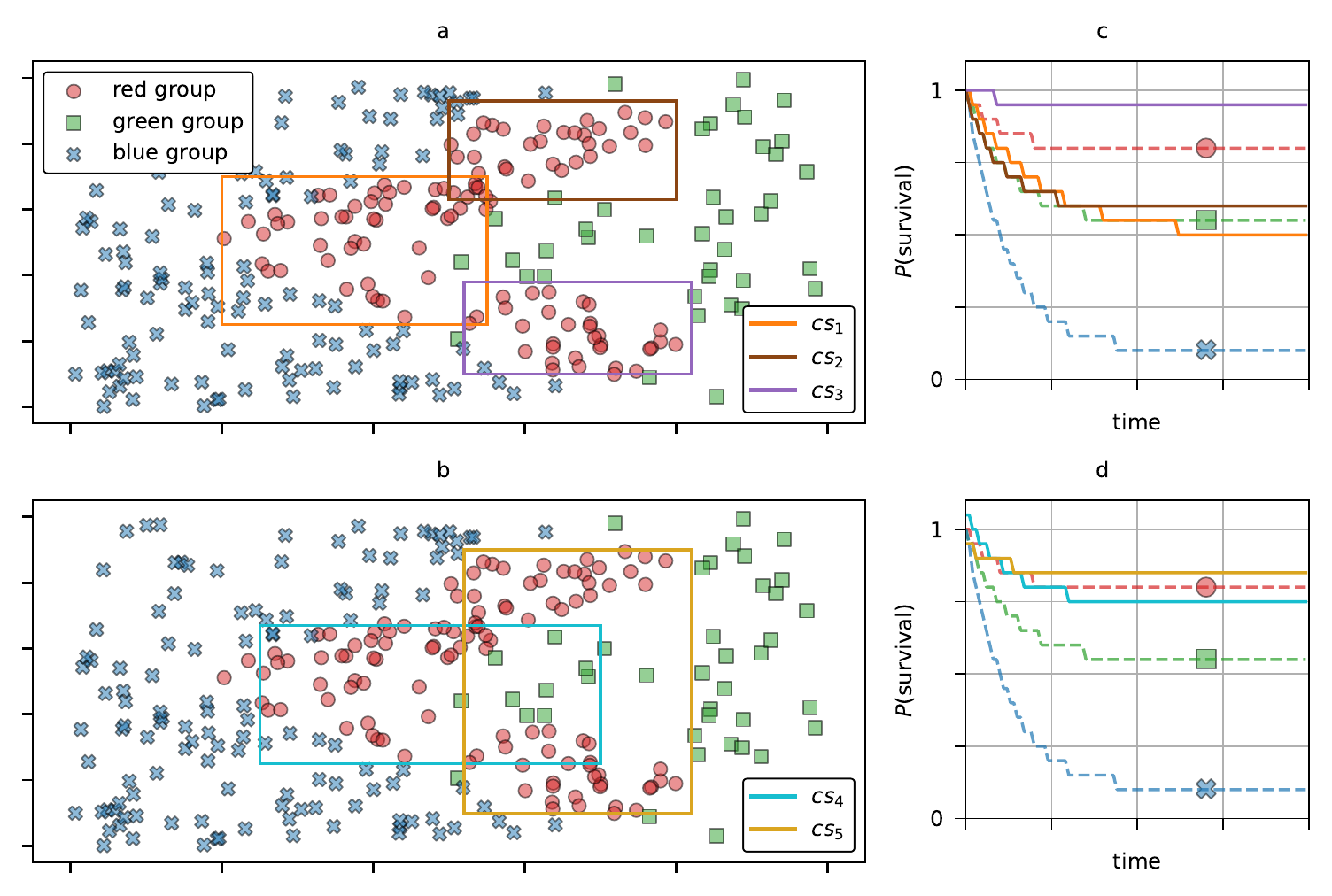}
	\caption{Example data set with three groups (red, green, and blue) visualized in a two dimensional space. (a) Classical contrast sets tend to differentiate the group of interest (red) from the remaining groups. (b) Survival contrast sets aim at resembling the survival function estimate of the entire group. Therefore, classical contrast sets $cs_1$, $cs_2$, and $cs_3$ whose survival characteristic is different to that of the entire group (c) were replaced with $cs_4$ and $cs_5$ which are more consistent with the group estimate (d).}
	\label{fig:illustration}
\end{figure}

In this paper, we present \algo{}, an algorithm for contrast set (CS) mining based on the sequential covering heuristic. Our approach follows the observation that contrast set mining is a special case of classification rule learning with particular emphasis put on maximizing the support difference between groups~\citep{webb2003detecting} (see Figure~\ref{fig:illustration}a for an intuitive example). 

While sequential covering has been previously employed for contrast set generation~\citep{novak2009csm},~\algo{} introduces numerous novelties. Multiple passes accompanied with attribute penalization scheme allow generating contrast sets describing same examples with different attributes -- the feature not ensured by the standard sequential covering, though crucial for the knowledge discovery potential. Additionally, instead of optimizing weighted relative accuracy,~\algo{} maximizes by default correlation between predicted and target variable which was shown to have better discriminating capabilities~\citep{wrobel2016rule}. 


Generalizing our algorithm for regression and survival data enables identifying contrast sets whose label attribute/survival prognosis is consistent with the label/prognosis for the predefined contrast groups. This ability, not provided by the existing methods, allows revealing new and interesting dependencies in the data (Figure~\ref{fig:illustration}b--d). 
 
The large-scale experiments on over 130 data sets from various areas allowed reliable investigation of algorithm properties and comparison with competing approaches. Detailed case studies performed on two selected medical sets further confirmed the usability of \algo{}. The algorithm was implemented in Java as a part of the RuleKit suite~\citep{gudys2020rulekit} available at GitHub under GNU AGPL~3 license (\link{https://github.com/adaa-polsl/RuleKit}). The versatility of the package (XML-based batch mode, R/Python package, RapidMiner plugin) greatly facilities application of \algo{} to real-life problems distinguishing it from many methods described in the literature that lack open implementations. 

The rest of the article is organized in four sections. Section 2 introduces related work. Section 3 describes in detail RuleKit-CS algorithm and demonstrates its operation on a synthetic example. The time complexity analysis is also provided there. Experimental verification of the presented method is given in Section 4. The study is concluded in Section 5.

\section{Related work}
The contrast set mining was formulated by Bay and Pazziani~\citep{bay2001detecting} as a problem of identifying differences between contrasting groups in multivariate data. Let $D$ indicate a set of examples (observations) described by conditional attributes $A = \{a_1, a_2,\ldots, a_m\}$ and assigned to one of the groups from $\{G_1, G_2,\ldots, G_n\}$. A contrast set was originally defined for data sets with categorical attributes only as a conjunction of attribute-value pairs with each attribute appearing at most once: $a_{i_1} = v_{i_1} \cand a_{i_2} = v_{i_2} \cand \ldots \cand a_{i_k} = v_{i_k}$, where $i_1, i_2, \cdots, i_k \in [1, \cdots, m]$. This definition can be straightforwardly generalized to continuous attributes by replacing attribute values with intervals. A support of a contrast set $cs$ in a group $G_i$ is calculated as the fraction of examples from $G_i$ covered by $cs$. The aim of contrast set mining is to find contrast sets with high support in the group of interest (here, referred to as \textit{positive}) and low support in the remaining groups (\textit{negative}). Contrast set requirements were formally defined~\citep{bay2001detecting} as:
\begin{linenomath}
\begin{equation}
\label{eq:cnd_siginificant}
P(cs | G_i) \neq P(cs | G_j),
\end{equation}
\begin{equation}
\label{eq:cnd_large}
|\Call{Support}{cs,G_i} - \Call{Support}{cs,G_j}| < \delta,
\end{equation}
\end{linenomath} 
with $\delta$ being a user-defined minimum support difference. Contrast sets fulfilling conditions~(\ref{eq:cnd_siginificant}) and~(\ref{eq:cnd_large}) are referred to as \textit{significant} and \textit{large}, respectively.

Bay and Pazziani proposed a contrast set mining algorithm STUCCO based on set-enumeration trees~\citep{bayardo1998efficiently}. The method investigates all potential contrast sets and selects those which (i) pass the statistical test of independence w.r.t. group membership and (ii) fulfill minimum support difference. Since the number of contrast sets grows exponentially with the number of attributes, the pruning strategies were incorporated to limit the search space. CIGAR~\citep{Hilderman05astatistically} extended the idea of STUCCO by introducing three additional constraints: (i) minimum support, (ii) minimum correlation, (iii) minimum correlation difference. 

Webb et al.~\citep{webb2003detecting} showed that contrast set mining is a special case of rule learning and confirmed that Magnum Opus, an implementation of a general-purpose association rule learner OPUS\_AR~\citep{webb2000}, was suitable for contrast set identification. Another analogy was shown by Krajl et al., who successfully applied subgroup discovery for mining contrast sets in various brain conditions~\citep{kralj2007brain, kralj2007diseases}. The equivalence between contrast set mining, subgroup discovery, and identifying emerging patterns was formally shown in~\citep{novak2009supervised}. The idea was followed in CSM-SD~\citep{novak2009csm} package which employed a subgroup discovery algorithm CN2-SD~\citep{lavrac2004} to the contrast set mining. As authors showed, the weighted relative accuracy (WRA)~\citep{Lavrac1999} used by CN2-SD corresponds to the support difference criterion from STUCCO. Though, unlike the latter, CN2-SD explores the search space using sequential covering. Other subgroup discovery algorithms, like pysubgroup~\cite{lemmerich2018pysubgroup} are also suitable to the identification of contrast sets. The detailed review of subgroup discovery approaches with their properties were presented by Atzmueller~\citep{Atzmueller2015}.  

Alternative approaches to contrast set mining include COSINE~\citep{simeon2011cosine},\linebreak{}DIFF~\citep{liu2014tree},  SciCSM~\citep{zhu2015scicsm}, or Exceptional Contrast Set Mining~\citep{nguyen2016exceptional}. There were also attempts to generate contrast sets in temporal data~\citep{magalhaes2015contrast} or to introduce fuzziness to contrast-based models~\citep{ahmed2022fuzzy}. 

While the majority of contrast set research concerned medical data~\citep{kralj2007brain, kralj2007diseases, novak2009csm, ahmed2022fuzzy}, the different areas of application like aircraft incidents~\citep{nazeri2008contrast}, software crashes~\citep{qian2020debugging}, or folk music~\citep{Neubarth2016} were also investigated in the literature.

\section{The algorithm}


\subsection{Separate and conquer contrast sets learning}

As presented in~\citep{webb2003detecting}, contrast set mining can be considered as a special case of classification rule learning with group $G$ being a label attribute. This idea is followed by our algorithm which employs sequential covering -- a well established method for rule induction. By selecting appropriate quality measure which controls learning process and introducing support constraints, the heuristic was suited for discovering contrast sets. 

The general idea of separate and conquer is an iterative addition of rules to the initially empty set as long as all positive examples become covered. To enable generation of many contrast sets describing the same subset of examples with different attributes, \algo{} performs multiple sequential covering passes. Contrast set redundancy between passes is prevented by using attribute penalization mechanism and/or different coverage requirements. By default, contrast sets are established in \textit{one vs all} scheme, i.e., the examples from the investigated group are differentiated from all the others. Optionally, the induction in \textit{one vs one} variant can be performed. In this mode, a single group is considered negative, while examples from the remaining groups are discarded. 

Given a group of interest $G_i$, let $D^{\oplus}$ and $D^{\ominus}$ indicate subsets of, respectively, positive and negative examples from $D$. By denoting a subset of examples from $D$ covered by a contrast set $cs$ as $\Call{Cov}{cs, D}$, we can define the elements of a $cs$ confusion matrix as: $P = |D^{\oplus}|$, $N = |D^{\ominus}|$, $p = |\Call{Cov}{cs, D^{\oplus}}|$, $n = |\Call{Cov}{cs, D^{\ominus}}|$. 
Additionally, let $D_U^{\oplus}$ be a subset of yet uncovered positive examples and $|\Call{Cov}{cs, D_U^{\oplus}}| = p_\textrm{new}$.

\begin{algorithm}[!t]
	\footnotesize
	\begin{algorithmic}[1]
		\caption{Separate-and-conquer contrast set induction.}
		\label{alg:conquer}
		\Require
		$D$---data set consisting of positive ($D^{\oplus}$) and negative ($D^{\ominus}$) examples, 
		\MINCOVall---sequence of investigated values of \mincovall{} (minimum positive support), 
		\mincov--minimum positive support for yet uncovered examples,
		\maxneg---maximum ratio of negative to positive support,
		\maxpass---maximum number of sequential covering passes for a single $\mincovall$,
		\qual---quality measure that drives learning process
		\Ensure $CS$---contrast sets.
		\State $CS \gets \emptyset$ \Comment{start from an empty contrast set collection}
		\For{$\mincovall \IN \MINCOVall$}
		\State $\Call{ResetAttributePenalties}{}$
		\For{$\textit{pass}~\textbf{in}~1 \ldots \maxpass$} \Comment{multiple passes}
		\State $CS_{pass} \gets \emptyset$ \Comment{contrast sets for current pass}
		\State $D_{U}^{\oplus} \gets D^{\oplus}$	\Comment{set of uncovered positives}
		\Repeat
		\State $cs \gets \Call{Grow}{D, D_{U}^{\oplus}, \mincovall, \mincov, \maxneg, \qual}$ 
		\If{$cs \neq \emptyset$}
		\State $cs \gets \Call{Prune}{cs,D,\maxneg,\qual}$ 
		\State $CS_{pass} \gets CS_{pass} \cup \{cs\}$ \Comment{add a contrast set}
		\State $D_{U}^{\oplus} \gets D_{U}^{\oplus}\setminus\Call{Cov}{cs, D_U^{\oplus}}$ \Comment{update set of uncovered examples}
		\State $\Call{UpdateAttributePenalties}{cs}$
		\State $\Call{CalculateRedundancy}{cs}$ 
		\EndIf	
		\Until{$cs = \emptyset$}	\Comment{end current pass}
		\If{$CS_{pass} \setminus CS = \emptyset$} \Comment{break if no new contrast sets discovered}
		\State \textbf{break} 
		\EndIf
		\State $CS \gets CS \cup CS_{pass}$ \Comment{add contrast sets from the current pass}
		\EndFor
		\EndFor
		\State \Return{$CS$}
	\end{algorithmic}
\end{algorithm}

The pseudocode of the contrast set mining for a group $G_i$ is presented in Algorithm~\ref{alg:conquer}. The procedure is controlled by the following parameters (the default values were established experimentally and are given here for convenience):
\begin{itemize}
\item $\mincovall$ --- a minimum positive support of a contrast set ($p / P$). By repeating the induction for decreasing $\mincovall$ and combining the results, the algorithm renders contrast sets from most general to most specific. This gives a broad view of the data at the beginning and provides more detailed dependencies later on. The default sequence of investigated $\mincovall$ values was established experimentally as $\MINCOVall = (0.8, 0.5, 0.2, 0.1)$. 
 
\item $\mincov$ --- a minimum positive support of a contrast set calculated using previously uncovered examples ($p_\textrm{new} / P$; 0.1 by default). The parameter ensures the convergence of a separate and conquer pass and is an equivalent of \textit{mincov} in our algorithm for rule induction~\citep{gudys2020rulekit}. Note, that $\mincov$ also determines the stop condition -- when the fraction of uncovered positives falls below its value, no more contrast sets fulfilling the requirement could be generated and the pass ends.  
    
\item $\maxneg$ --- a maximum ratio of negative to positive supports\linebreak($(n \cdot P)/(p \cdot N)$; 0.5 by default). The parameter ensures the basic requirement of the contrast set, i.e., the high support in the group of interest and low in the remaining ones.
\item $\maxpass$ --- a maximum number of sequential covering passes for a single $\mincovall$ (5 by default). 

\item $\qual$ --- rule quality measure that drives the learning process. The measure is expressed as a function of confusion matrix elements\linebreak$(p, n, P, N)$ and allows balancing support and precision of resulting contrast sets. Many rule quality measures with various characteristics have been defined~\citep{kamber1996evaluating, hilderman2013knowledge, tan2002selecting, greco2004can}. One of them is the correlation between predicted and target variables defined as:
\begin{linenomath}
\begin{equation}
\textrm{Correlation} = \frac{p \cdot N-P \cdot n}{\sqrt{P \cdot N \cdot (p+n) \cdot (P-p+N-n)}}.
\end{equation}
\end{linenomath}
Due to its valuable properties, the measure was employed in rule induction, subgroup discovery, or evaluation of association rules~\citep{xiong2004exploiting, geng2006interestingness, janssen2010quest}. In particular, the correlation is monotonic (increasing in $p$ for fixed $n$, decreasing in $n$ for fixed $p$), symmetric (if we negate the premise or the consequence, the correlation value becomes the additive inverse), and takes values from the $[-1,1]$ interval. Additionally, it belongs to the confirmation measures~\citep{eells2002symmetries} -- it is positive if the contrast set precision $p/(p+n)$ exceeds the a priori group precision $P/(P+N)$, and negative otherwise. It has been also shown that sorting a sequence of rules w.r.t. Correlation renders very similar ordering as weighted relative accuracy (average Kendall rank correlation coefficient over 0.8)~\citep{sikora2012wybrane}.
Finally, we experimentally confirmed Correlation to be a convenient support-precision trade off~\citep{wrobel2016rule,sikora2013data}, thus we set it as a default quality measure for traditional contrast sets 


Note, that a different quality measure is used for regression/survival data where the procedure aims at rendering contrast sets which are consistent with the entire group w.r.t. the label attribute/survival prognosis (see Subsection~\ref{subsec:reg-surv} for details).   	
\end{itemize}

The function calls in lines 3, 13, and 14 of Algorithm~\ref{alg:conquer} belong to the attribute penalization mechanism which is described in details in Subsection~\ref{subsec:penalization}.

An induction of a contrast set consists of two steps: growing and pruning. As shown in Algorithm~\ref{alg:grow}, the former starts from an empty premise and adds conditions iteratively, each time selecting the one optimizing $\qual$ (correlation for classical contrast sets, label consistency for regression, survival function consistency for survival problems; lines 9--11). Conditions, which cause the contrast set to violate $\mincovall$ or $\mincov$ constraints are discarded (line 8). Growing stops when there are no more conditions fulfilling the support requirements. As in principle, a quality measure used for condition evaluation rewards $p$ and penalizes $n$, the ratio of negative to positive supports, $(n \cdot P) / (p \cdot N)$ decreases during growing stage. Therefore, the $\maxneg$ requirement is verified for the fully grown contrast set (line 14).

\begin{algorithm}[!p]
	\footnotesize
	\begin{algorithmic}[1]
		\caption{Growing a contrast set.}
		\label{alg:grow}
		\Require
		$D = D^{\oplus} \cup D^{\ominus}$---training dataset,
		$D_{U}^{\oplus}$---set of uncovered positive examples,
		\mincovall---minimum positive support, 
		\mincov---minimum positive support for yet uncovered examples,
		\maxneg---maximum ratio of negative to positive support,
		\qual---quality measure that drives learning process
		\Ensure
		$cs$---grown contrast set.
		
		\Function{Grow}{$D$, $D_U^{\oplus}$, $\mincovall$, $\mincov$, $\maxneg$, $\qual$}
		
		\State $cs \gets \emptyset$ \Comment{start from an empty premise}
		\Repeat \Comment{iteratively add conditions}
		\State $c_\textrm{best} \gets \emptyset$ \Comment{current best condition}
		\State $q_\textrm{best} \gets -\infty,\quad \textrm{cov}_\textrm{best} \gets -\infty$ \Comment{best quality and coverage}
		
		
		\For{$c \IN \Call{PossibleConditions}{\Call{Cov}{cs, D}}$}
		\State $cs' \gets cs \cand c$ \Comment{extend contrast set}
		\LeftComment*{4.5em}{verify support constraints}
		\If {$\frac{|\Call{Cov}{cs',D^{\oplus}}|}{|D^{\oplus}|} \geq \mincovall \AND \frac{|\Call{Cov}{cs',D_U^{\oplus}}|}{|D^{\oplus}|} \geq \mincov}$ 
		\State $q \gets \Call{Evaluate}{cs, D, \qual}$ 
		
		\If {$q > q_\textrm{best}$ \OR ($q = q_\textrm{best}$ \AND $|\Call{Cov}{cs',D}| > \textrm{cov}_\textrm{best}$)}
		\State $c_\textrm{best} \gets c,\quad q_\textrm{best} \gets q$,\quad $\textrm{cov}_\textrm{best} \gets |\Call{Cov}{cs',D}|$
		\EndIf
		
		\EndIf
		\EndFor			
		\State $cs \gets cs \cand c_\textrm{best}$ \Comment{extend contrast set with best condition}
		\Until{$c_\textrm{best} = \emptyset$}
		\LeftComment*{1.5em}{verify negative to positive support ratio}
		\If{$cs \neq \emptyset \AND \frac{|\Call{Cov}{cs, D^{\ominus}}|}{|D^{\ominus}|}  \mathbin{/} \frac{|\Call{Cov}{cs, D^{\oplus}}|}{|D^{\oplus}|} \le \maxneg$} 
		\State \Return{$cs$}
		\Else 
		\State \Return{$\emptyset$}
		\EndIf
		\EndFunction
	\end{algorithmic}
\end{algorithm}

\begin{algorithm}[!p]
	\footnotesize
	\begin{algorithmic}[1]
		\caption{Pruning a contrast set.}
		\label{alg:prune}
		
		\Require
		$cs$---input contrast set,
		$D = D^{\oplus} \cup D^{\ominus}$---training dataset,
		\mincovall---minimum positive support, 
		\qual---quality measure that drives learning process
		\Ensure $cs$---pruned contrast set (algorithm operates in-place)
		
		\Function{Prune}{$cs$, $D$, $\mincovall$, $\maxneg$, $\qual$}
		
		\Repeat
		\State $c_\textrm{remove} \gets \emptyset$ \Comment{condition to remove}	
		\State $q_\textrm{best} \gets \Call{Evaluate}{cs, D, \qual}$
		
		\For{$c \IN cs$} \Comment{iterate over all conditions}
		\State $cs' \gets cs \setminus c$ \Comment{try to remove a condition}
		\If{$\frac{|\Call{Cov}{cs', D^{\ominus}}|}{|D^{\ominus}|}  \mathbin{/} \frac{|\Call{Cov}{cs', D^{\oplus}}|}{|D^{\oplus}|} \le \maxneg$} \Comment{verify constraint}
		\State $q \gets \Call{Evaluate}{cs, D, \qual}$
		\If{$q \ge q_\textrm{best}$}
		\State $c_\textrm{remove} \gets c, \quad q_\textrm{best} \gets q$
		\EndIf
		\EndIf
		\EndFor
		
		\State $cs \gets cs \setminus c_\textrm{remove}$ 
		
		\Until{$c_\textrm{remove} = \emptyset \OR |cs| = 1$} \Comment{no conditions to remove or single condition left}
		
		\State \Return{$cs$}
		\EndFunction
	\end{algorithmic}
\end{algorithm}

If growing produces an empty contrast set (no conditions fulfilling\linebreak{}$\mincov$ or $\mincovall$) or $\maxneg$ constraint is violated, the contrast set is discarded and the current separate and conquer pass ends. Otherwise, the pruning starts (Algorithm~\ref{alg:prune}). The procedure, as an opposite to growing, removes conditions from the premise, each time making an elimination leading to the largest improvement in the rule quality (lines 8--10), with a restriction that $\maxneg$ requirement cannot be violated (line 7). The iteration stops when no such eliminations further exist. 
	
Note, that in the original contrast set definition~\citep{bay2001detecting}, the difference between supports $(p/P - n/N)$ was controlled rather then their ratio. This, however, would filter out contrast sets with very good discriminating capabilities but moderate support. For instance, if we set the minimum support difference to 30\%, the contrast set that covers 80\% of positives and 50\% of negatives would be accepted, but the one covering 25\% of positives and no negatives not. As we believe the latter is also an interesting contrast set, we decided to control the ratios of supports with $\maxneg$ instead of their difference. This way, the required support difference becomes dependent on the support value itself (which is controlled by $\mincovall$).     

Unlike many existing data mining algorithms, RuleKit-CS intrinsically manages numerical attributes by investigating all possible splits. Missing values are also handled -- an observation lacking value of an attribute is considered as not fulfilling the condition built upon it. These features are important from the applicability perspective as they reduce the number of preprocessing steps required to run the analysis.

\subsection{Regression and survival data}
\label{subsec:reg-surv}
The problem for contrast set mining can be generalized for regression and survival data sets. For this purpose, we assume the presence of the label variable $L$ apart from the group $G$. For regression, the label is continuous, while in survival problems, it represents a binary censoring status with 0 and 1 representing censored (event-free) and non-censored (event-subjected) observations, respectively. The status variable is accompanied with a survival time $T$, i.e., the time of the observation for event-free examples or the time before the occurrence of an event. 

The analysis of regression and survival data sets by \algo{} differs from the classical contrast set mining based only on the group attribute. Instead of directly diversifying supports across groups by optimizing Correlation measure, our algorithm takes into account the label. In particular, for regression problems it establishes mean labels of (i) examples covered by a contrast set $cs$ and (ii) the entire group of interest. The quality measure to be maximized (Algorithm~\ref{alg:grow}, line 9; Algorithm~\ref{alg:prune}, line 8) is then defined as an opposite of an absolute difference of these two means: 
\begin{linenomath}
\begin{equation}
\Call{Evaluate}{cs,D} = -\left| 
	\frac{\sum_{x_i \in \Call{Cov}{cs,D}}{L_i}}{|\Call{Cov}{cs,D}|} -
	\frac{\sum_{x_i \in D^\oplus}{L_i}}{|D^\oplus|}
\right|.
\end{equation}
\end{linenomath}
Consequently, the algorithm tends to extract contrast sets with label consistent with a label of the entire group.

Survival data sets are handled analogously, but instead of taking label $L$ as an outcome, the algorithms considers Kaplan-Meier survival function estimates~\citep{kaplan1958} of examples covered by a contrast set and the entire group. Since the aim is to minimize the difference between these two survival prognosis, the algorithm maximizes an opposite of the log-rank test statistics:
\begin{linenomath}
\begin{equation}
\Call{Evaluate}{cs, D} = -\Call{LogRank}{\Call{Cov}{cs,D},D^\oplus}.
\end{equation}
\end{linenomath}

\subsection{Contrast set diversity}
\label{subsec:penalization}
The crucial feature of contrast sets are their descriptive capabilities, i.e., the ability to represent interesting and unknown relationships between conditional attributes and a group label. In particular, we want different contrast sets to represent different concepts in the attribute space. Consequently, contrast sets which, at the same time, contain similar attributes and cover similar group of examples as previously generated contrast sets can be considered redundant and should be avoided. As classical separate and conquer algorithm aims at maximizing quality measure discarding the aforementioned aspects, we introduce a novel heuristic which prevents redundancy in the generated contrasts sets. 

The mechanism consists of two components: (i) the penalty $\penal$ for reusing already utilized attributes, which can be compensated by (ii) the reward $\reward$ for covering previously uncovered examples. The components are incorporated into quality evaluation of the contrast set candidate $cs_k$ according to the formula:
\begin{linenomath}
\begin{equation}
q' = q (1 - s\penal) \reward
\end{equation}
\end{linenomath}        
with $q$ and $q'$ representing the input and the modified quality, respectively, and $s \geq 0$ being the penalty strength. The penalty $\penal$ for the current contrast set $cs_k$ reflects to what extent the attributes employed by contrast sets $cs_1, \ldots, cs_{k-1}$ are reused in $cs_k$. 
For each attribute $a \in A$ we define an attribute penalty $\penal_a$ as
\begin{linenomath}
\begin{equation}
\pi_a = \frac{\sum_{i=1}^{k-1} \Call{Contains}{cs_i, a}}{\sum_{a_j \in A} \sum_{i=1}^{k-1} \Call{Contains}{cs_i, a}}
\end{equation}
\end{linenomath} 
with $\Call{Contains}{cs, a}$ returning $1$ if contrast set $cs$ contains attribute $a$ and $0$ otherwise. The penalty for contrast set $cs_k$ is a sum of attribute penalties for all attributes contained in $cs_k$.

For instance, let us assume that there are four conditional attributes $a_1, \ldots, a_4$. After inducing two contrast sets $cs_1$ and $cs_2$ containing attributes $ \{a_2, a_4\}$ and $\{a_1, a_2\}$, respectively, the attribute penalties equal to $\penal_{a_1}=1/4$, $\penal_{a_2}=2/4$,  $\penal_{a_3}=0$, $\penal_{a_4}=1/4$. The contrast set candidate $cs_3$ built upon attributes $\{a_2, a_3, a_4\}$ would be penalized with $\penal= \penal_{a_2} + \penal_{a_3} + \penal_{a_4} = 3/4$. The proposed penalization scheme has a property of the tabu search. With a small number of already used attributes at the beginning, there is a strong pressure for selecting different features. This results in less redundant and, potentially, more interesting contrast sets. As consecutive contrast sets are induced, the attribute penalties become more even reducing the $\penal$ effect and allowing algorithm to cover remaining positive examples. 

The penalty alone does not take into account the fact that a contrast set built upon already used attributes may still be interesting as long as it covers previously uncovered examples. For this purpose, the reward $\reward$ was introduced. The value of $\reward$ depends on the contribution $p_\textrm{new} / p$ of previously uncovered positive examples in all positives covered by the contrast set. The rewards decreases linearly from $1 / (1 - s\penal)$ (full penalty compensation) when the contrast set covers only new examples ($p_\textrm{new} / p = 1$) up to $1$ (no compensation) at some boundary value of $p_\textrm{new} / p$. Note, that the proposed penalty-reward scheme does not allow modified quality $q'$ to exceed the initial $q$ value. The penalization procedure is incorporated into both, growing and pruning stages and considers cases of multiple occurrences of an attribute in a contrast set. E.g., if $cs_k$ candidate contains condition $a > x$, an attempt to close the interval by adding $a < y$ does not affect $\penal$ value as $\penal_a$ component has been already included.   

In \algo{}, the attribute penalties are updated after induction of a contrast set (Algorithm~\ref{alg:conquer}, line 13) and are reset for every investigated value of $\mincovall$ (Algorithm~\ref{alg:conquer}, line 3). Consequently, the procedure does not prevent from inducing similar contrast sets across different $\mincovall$ values. Therefore, at the very end of the induction, we quantify the redundancy of every contrast set (Algorithm~\ref{alg:conquer}, line 14) which can be further used for selecting most interesting dependencies. Let us define a similarity between contrast sets $c_i$ and $c_j$ as:
\begin{linenomath}
\begin{equation}
\begin{split}
\label{eq:similarity}
\Call{Similarity}{cs_i, cs_j} & = J\left(\Call{Attr}{cs_i},\Call{Attr}{cs_j}\right) \\ 
					& \times J\left(\Call{Cov}{cs_i,D^{\oplus}},\Call{Cov}{cs_j,D^{\oplus}}\right)
\end{split}
\end{equation}
\end{linenomath}
with $J$ being the Jaccard index and $\Call{Attr}{cs}$ representing the set of $cs$ attributes. The redundancy of contrast set $cs_k$ is defined as a similarity to its most similar predecessor:
\begin{linenomath}
\begin{equation}
\label{eq:redundancy}
\Call{Redundancy}{cs_k} = \textrm{max}_{i=1}^{k-1} \left( \Call{Similarity}{cs_k, cs_i} \right)
\end{equation} 
\end{linenomath}

\subsection{Synthetic example} 
\begin{figure}
	\centering
	\includegraphics[width=1.0\textwidth]{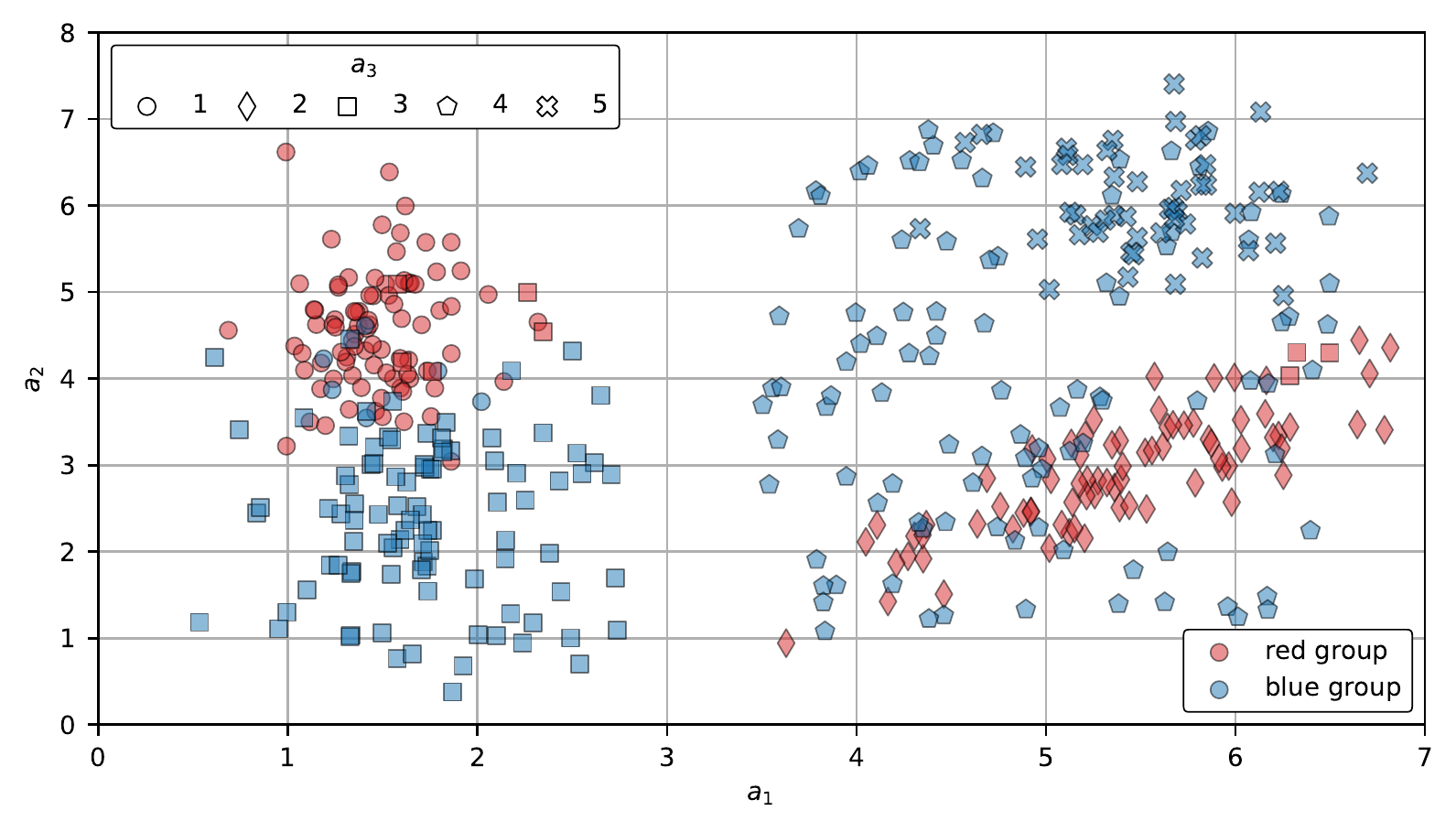}
	\caption{Synthetic data set with 320 examples from two groups (\ii{red} and \ii{blue}) described by two numerical ($a_1$ and $a_2$) and one categorical ($a_3$) attribute.}
	\label{fig:synth}
\end{figure}

The synthetic data set consists of 320 examples from two groups (170~\textit{red}, 250~\textit{blue}) described by two numerical ($a_1$, $a_2$) and one categorical ($a_3$) attribute. As presented in Figure~\ref{fig:synth}, the elements from group \textit{red}, which was selected as a group of interest, are arranged in two clusters in the attribute space. The left cluster is almost perfectly separable from blue examples by $a_3$, and well separable when using $a_1$ and $a_2$ together. The right cluster is perfectly separable with $a_3$ and poorly separable when using $a_1$ and/or $a_2$. All the analyses presented below were performed for $\mincovall = 0.1$ which is the smallest among values investigated by \algo{} -- the algorithm aggregates results from $\mincovall \in (0.8, 0.5, 0.2, 0.1)$.

As a preliminary step, we applied on a data set a single separate and conquer pass without attribute penalties, as in the classification rule induction. This rendered the following results for \ii{red} group (the numbers of positive and negative examples are given on the right):
\begin{description}
\setlength{\itemsep}{0pt}%
\item[cs-1:] {\small $a_3 = 2 \hfill p=77, n=0$}
\item[cs-2:] {\small $a_3 = 1 \hfill p=85, n=6$}
\end{description}

The contrast sets describe right (cs-1) and left (cs-2) \ii{red} clusters with a use of $a_3$ nominal attribute and are characterized by very good quality. Since the number of uncovered examples equals to $8$ which is lower than $\mincov \times P = 17$, the sequential covering stopped at this point without attempt to cover remaining instances. The contrast set describing left cluster with $a_1$ and $a_2$ numerical attributes remained undiscovered, which was expected as only one sequential covering pass was run.

In order to learn more contrast sets, we applied 5 covering passes with attribute penalization to ensure the CS diversity. The penalty strength $s$ was set to $0.5$, while the reward for covering different examples was disabled. The \ii{red} group was described by the contrast sets below (for convenience, we provide pass numbers for each CS):

\begin{description}
	\setlength{\itemsep}{0pt}%
	\item[cs-3:] {\small $a_3 = 2 \hfill p=77, n=0, \textrm{pass} = 1$}
	\item[cs-4:] {\small $a_1 \in [0.92, 1.78) \cand a_2 \in [3.55, \infty) \hfill p=73, n=6, \textrm{pass} = 1$}
	\item[cs-5:] {\small $a_2 \in [2.15, 5.61) \cand a_3 \neq 4 \hfill p=157, n=69, \textrm{pass} = 1$}
	\item[{\color{Gray}cs-6}:] {\color{Gray}\small $a_3 = 2 \hfill p=77, n=0, \textrm{pass} = 2$}
	\item[cs-7:] {\small $a_3 = 1 \hfill p=85, n=6, \textrm{pass} = 2$}
\end{description}
   
As previously, the first (highest quality) contrast set is the one describing the right cluster with $a_3$. The second best candidate was $a_3 = 1$. However, as $a_3$ was the only already used attribute, its penalty equaled to $\pi_{a_3} = 1/1$ resulting in the quality modifier $(1 - s\pi_{a_3})$ of 0.5. Therefore, the algorithm preferred cs-4 describing left cluster with a use of $a_1$ and $a_2$. Contrast sets cs-3 and cs-4 left 20 uncovered red examples. As this was more then $\mincov \times P$, an attempt was made to cover the remaining examples with cs-5. This contrast set describes entire left cluster and substantial part of the right cluster, but at the same time it covers large number of negative examples, thus it is characterized by low quality. After that, the second sequential covering pass started. The penalties related to the attributes were equal to $\pi_{a_1} = 1/5$, $\pi_{a_2} = 2/5$, $\pi_{a_3} = 2/5$. This leaded to the contrast set cs-6 --- the duplicate of cs-3. This contrast set was filtered out in the post processing step. However, in order to reduce the chance of generating the same CS later on, it contributed to the attribute penalty. The next candidate was, as in the first pass, $a_3 = 1$. Due to lower penalty $\pi_{a_3} = 3/6$ the contrast set was accepted finishing the second sequential covering pass. The third pass did not introduce any novel contrast sets, fulfilling the stop condition.   

Eventually, the multiple sequential covering passes together with attribute penalties allowed discovering all interesting contrast sets. However, discarding the information about examples covered by CS in the penalization leads to the undesired situation where very good sets (cs-7) are learned after those of low quality (cs-5). To prevent this, we enabled rewards (with saturation set at $p_\textrm{new} / p = 0.2$) which resulted in the following contrast sets:

\begin{description}
	\setlength{\itemsep}{0pt}%
	\item[cs-8:] {\small $a_3 = 2 \hfill p=77, n=0, \textrm{pass} = 1$}.
	\item[cs-9:] {\small $a_3 = 1 \hfill p=85, n=6, \textrm{pass} = 1$}.
	\item[cs-10:] {\small $a_1 \in [0.92, 1.78) \cand a_2 \in [3.55, \infty) \hfill p=73, n=6, \textrm{pass} = 2$}.
	\item[{\color{Gray}cs-11}:] {\color{Gray}\small $a_3 = 2 \hfill p=77, n=0, \textrm{pass} = 2$}.
	\item[cs-12:] {\small $a_2 \in [2.15, 5.61) \cand a_3 \neq 4 \hfill p=157, n=69, \textrm{pass} = 2$}.
\end{description}
As cs-9 covered different examples then cs-8, the penalty for reusing $a_3$ was entirely compensated by reward. The second pass started from the last interesting contrast set cs-10 based on $a_1$ and $a_2$ attributes. This was followed by cs-8 duplicate and the low quality contrast set cs-12 (same as cs-5) induced in order to cover remaining \ii{red} examples.

Clearly, the fusion of multiple sequential covering passes combined with attribute penalties and rewards rendered the most convenient list of contrasts sets. Therefore, all these mechanisms are by default enabled in the presented algorithm. One must keep in mind though, that rewards are effectively working only in the first sequential covering pass as few uncovered observations (at most $\mincov$) are left for the following passes.

\subsection{Time complexity analysis}
Let us consider the time complexity of inducing a single contrast set. In the growing stage, conditions are added iteratively to the initially empty premise. Every addition decreases the contrast set support by at least one example resulting in at most $(1-\mincovall)|D|$ grown conditions (multiple conditions built upon same attribute are allowed). Every addition requires analyzing all possible split points on all attributes. In the worst case, all attributes are numerical and their values are unique. This equals to $2(|D|-1)|A|)$ candidates (two per split point) for an empty premise and decreases as the consecutive conditions are added. Eventually, the number of condition evaluations in growing is
\begin{linenomath}
\begin{equation}
\label{eq:complexity_growing}
\mathcal{O}\Big((1-\mincovall)|D|^2|A|\Big).
\end{equation}
\end{linenomath}
In the pruning stage, the algorithm removes conditions iteratively from the premise, every time making an elimination resulting in the highest quality value. In the worst case, the procedure retains only one from at most $(1-\mincovall)|D|$ conditions, leading to
\begin{linenomath}
\begin{equation} 
\mathcal{O}\Big((1-\mincovall)|D|^2\Big)
\end{equation}
\end{linenomath}
rule quality evaluations. Eventually, the time complexity of inducing a contrast set is given by Equation~\ref{eq:complexity_growing}. 

Every induced contrast set has to cover at least $\mincov$ yet uncovered examples resulting in no more than $\lfloor{\mincov^{-1}}\rfloor$ contrast sets. As the algorithm aggregates results for multiple $\mincovall$ performing at most $\maxpass$ passes for each investigated value, the final time complexity of RuleKit-CS is
\begin{linenomath}   
\begin{equation}
\label{eq:complexity_total}
\mathcal{O}\left(\sum_{\mincovall}(1-\mincovall) \cdot \maxpass \cdot \lfloor{\mincov^{-1}}\rfloor \cdot |D|^2|A|\right). 
\end{equation}
\end{linenomath}
Under reasonable assumptions that there are few $\mincovall$ values spread across $[0,1]$ range, this can be reduced to
\begin{linenomath}
\begin{equation}
\label{eq:complexity_reduced}
\mathcal{O}\Big(\maxpass \cdot \lfloor{\mincov^{-1}}\rfloor \cdot |D|^2|A|\Big) 
\end{equation}
\end{linenomath}
with condition evaluation being the basic operation.

The analysis of a condition requires (i) calculating rule quality and (ii) applying attribute penalty. The complexity of the former depends on the problem. For classification and regression, it is done in constant time, as $p$ and $n$ elements as well as average labels can be updated immediately while investigating consecutive split points. Evaluation in survival problems it is done in $\mathcal{O}(|D|)$ time, as it requires determining a survival estimates of covered examples and performing a log rank test. Applying attribute penalty is done in constant time as already employed attributes are kept in a hash table. 

As a result, a complexity from Equation~\ref{eq:complexity_reduced} increases by a factor $\mathcal{O}(|D|)$ when investigating survival data. In practice, the algorithm is sufficiently fast for moderately-sized data sets (i.e., thousands of instances described by a dozen or so features). This is due to several reasons: nominal attributes reducing the number of possible conditions, coverages represented as bit vectors for bit-parallel operations, performing less than $\maxpass$ due to lack of new contrast sets.  

\subsection{Implementation}

RuleKit-CS was implemented in Java as a part of RuleKit -- our versatile suite for rule-based learning~\citep{gudys2020rulekit} distributed at GitHub under GNU AGPL 3 license (\link{https://github.com/adaa-polsl/RuleKit}). Thanks to this, it is multi-platform and flexible -- it can be run in XML-configured batch mode, through R/Python package, or as a RapidMiner plugin. RuleKit-CS takes advantage of multi-core architecture of current central processors and is highly optimized (for instance, it employs bit-parallelism).  
\section{Experimental verification}
\subsection{Experimental setting}
The experiments were performed on 50 classification data sets downloaded from the UCI Machine Learning Repository~\citep{Dua2019} with class label being used as a group attribute. The sets represent problems from various domains and have different characteristics (size, dimensionality, types of attributes, missing values presence). The data is available at RuleKit repository as a collection of ARFF files (\link{https://github.com/adaa-polsl/RuleKit/tree/master/data/contrast-sets}). 
 
\algo{} was compared with well known contrast set induction algorithms: STUCCO~\cite{bay2001detecting}, CSM-SD~\cite{novak2009csm}, and pysubgroup~\citep{lemmerich2018pysubgroup}. As CSM-SD was shown by the authors to be an equivalent of CN2-SD subgroup discovery algorithm~\citep{lavrac2004}, we used an implementation of the latter from Orange 3~\citep{demvsar2013orange} package. Pysubgroup, in spite of being suited for subgroup discovery, can be used for contrast set mining due to comaptibility of these problems~\citep{novak2009supervised}.
For completeness, traditional classification rules induced by RuleKit suite~\citep{gudys2020rulekit} were also included in the analysis. The aforementioned algorithms, by rendering very different models, constituted a good reference for the experimental evaluation of \algo{}. 

CSM-SD and pysubgroup were executed with default settings (beam search was used in the latter), while RuleKit was configured so that parameters having their equivalents in \algo{} were set as in the latter (Correlation as a quality measure, $mincov=0.1$). This was to limit the effect of different parameters on the induction and investigate the novel algorithmic features introduced in the presented method. STUCCO was also run at its default configuration with an exception of \textit{kr-vs-kp} set, where \textit{maxorder} parameter was set to 6 due to computational reasons. Importantly, STUCCO, as the only method in the comparison, is limited to nominal features. Therefore, the analyses with this algorithm were preceded by a supervised entropy-based discretization of numerical attributes~\citep{Fayyad1992}. 

As \algo{} requires a negative support of a contrast set to be at most half of its positive support ($\maxneg = 0.5$), the results of the competing algorithms were subject to the analogous filtering. 
All the experiments were performed in \textit{one vs all} scheme (a group against all the others). Presented algorithm was run through RuleKit batch interface (XML files with experiments configuration can be found at the repository: \link{https://github.com/adaa-polsl/RuleKit/tree/master/examples/contrast-sets}).

\subsection{Comparison with other algorithms}

Executing multiple covering passes with attribute penalties is a crucial feature of \algo{} that make separate and conquer approach suitable for contrasts set induction. For this reason, as an initial step, we investigated the effect of attribute penalization in \algo{} on the background of the competitors.

\begin{figure}
\centering
\includegraphics{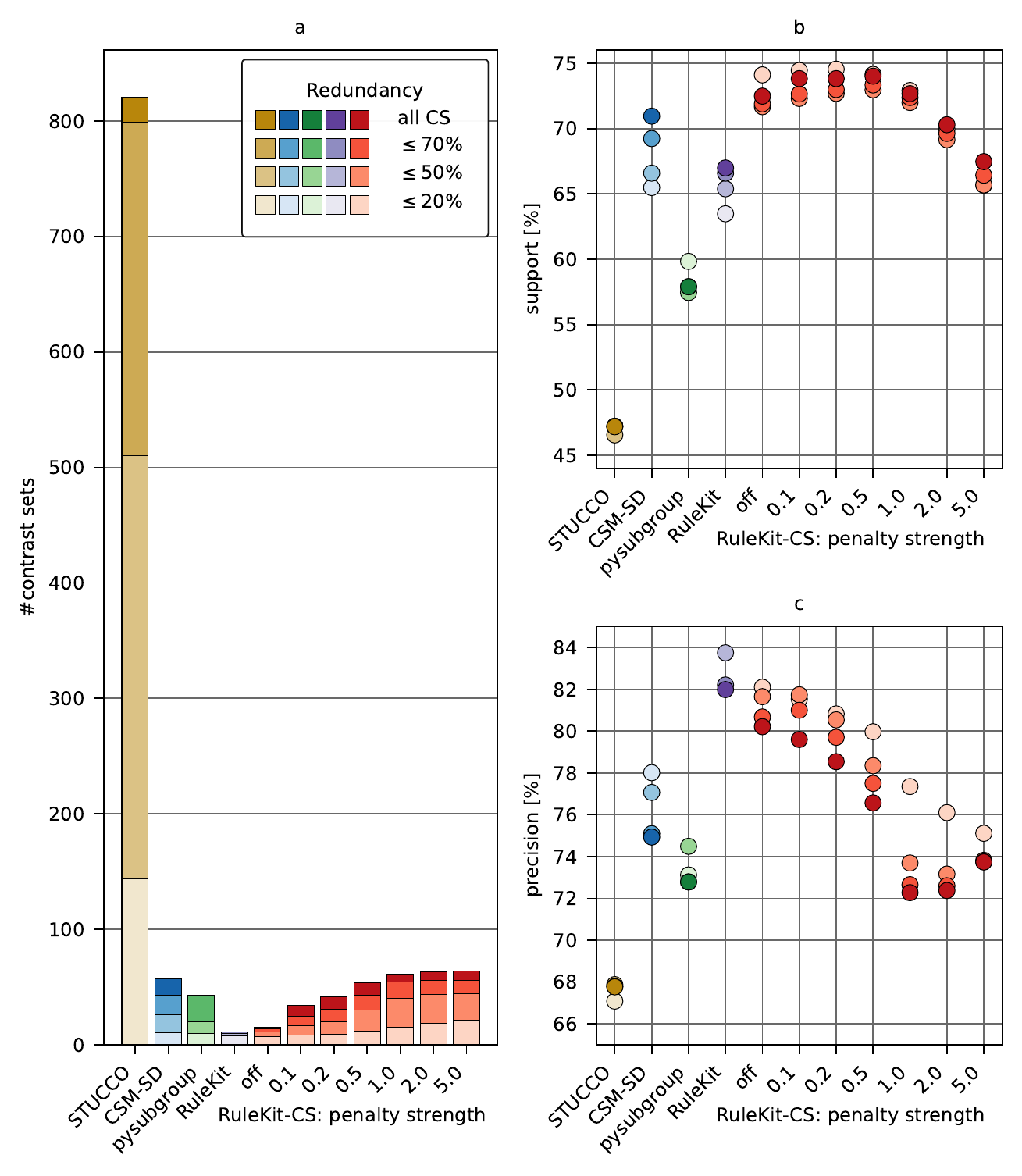}
\caption{The analysis of penalty strength in \algo{} against other contrast set induction algorithms: STUCCO, CSM-SD, pysubgroup, and RuleKit. The comparison includes the following performance indicators averaged over 50 data sets: (a) number of contrast sets, (b) average CS support, (c) average CS precision. The results for different contrast set redundancy thresholds (all CS, 70\%, 50\%, 20\%) are represented with increasing brightness. }
\label{fig:comparison}
\end{figure}

Figure~\ref{fig:comparison} presents algorithms' performance, i.e., number of contrasts sets, together with their average support and precision, summarized (averaged) over all 50 data sets. As the redundancy of individual contrast sets was established (Equation~\ref{eq:redundancy}), we show also the performance indicators of contrast sets not exceeding assumed redundancy thresholds: 70\%, 50\%, and 20\%. 

The most noticeable observation is that STUCCO, compared to its competitors, generated significantly more contrast sets, which at the same time were less general and less precise. This was probably due to fact that, by default, STUCCO mines contrast sets at small support difference with non-restrictive pruning thresholds. Therefore, CSM-SD, which generated less contrast sets with higher average support and precision (71\% and 75\%, respectively) was considered as a more convenient baseline. Pysubgroup produced fewer contrast sets, that were substantially less general (58\% support) and slightly less precise (73\%). Importantly, both methods generated CS from the entire redundancy range. The very different behaviour was observed for RuleKit, which produced several times less contrast sets than CSM-SD or pysubgroup, with almost no redundancy. This was due to fact that the algorithm performed a single sequential covering pass, thus the observations were usually covered by only one rule. While the support of the RuleKit contrast sets fell between CSM-SD and pysubgroup (67\%), the precision was the best among all investigated packages (82\%), which was expected for the algorithm specialized in the classification rule induction.         
 
The results of \algo{} strongly depended on the penalization scheme. When penalties were disabled, the algorithm rendered slightly more contrast sets than RuleKit. This was caused by aggregating results from four investigated values of $\mincovall$ parameter, i.e., 0.8, 0.5, 0.2, and 0.1, while RuleKit performed only one covering pass. The resulting contrast sets were characterized by superior support (above 72\%) and precision only slightly inferior to that of RuleKit (80\%). Running additional passes for each $\mincovall$ had no effect in this scenario, as due to lack of penalties all of them rendered duplicated contrast sets w.r.t. the initial pass. Increasing penalty strength $s$ enforced different induction paths in the consecutive covering passes for a given $\mincovall$. As a result, the average number of contrast sets increased and saturated for $s=1.0$ at the level slightly above CSM-SD. As for the support, for small penalties it was superior by a small margin to the non-penalty variant and started to deteriorate noticeably for $s \ge 0.5$. The different situation was in the case of precision which decreased consistently with the growing penalty. After the analysis of Figure~\ref{fig:comparison}, the penalty strength $s=0.5$ was selected as the best trade-off between the number of contrast sets (slightly lower than CSM-SD) and their quality (support and precision greater than CSM-SD by approx. 2 pp.).              

\begin{figure}
	\centering
	\includegraphics[scale=1]{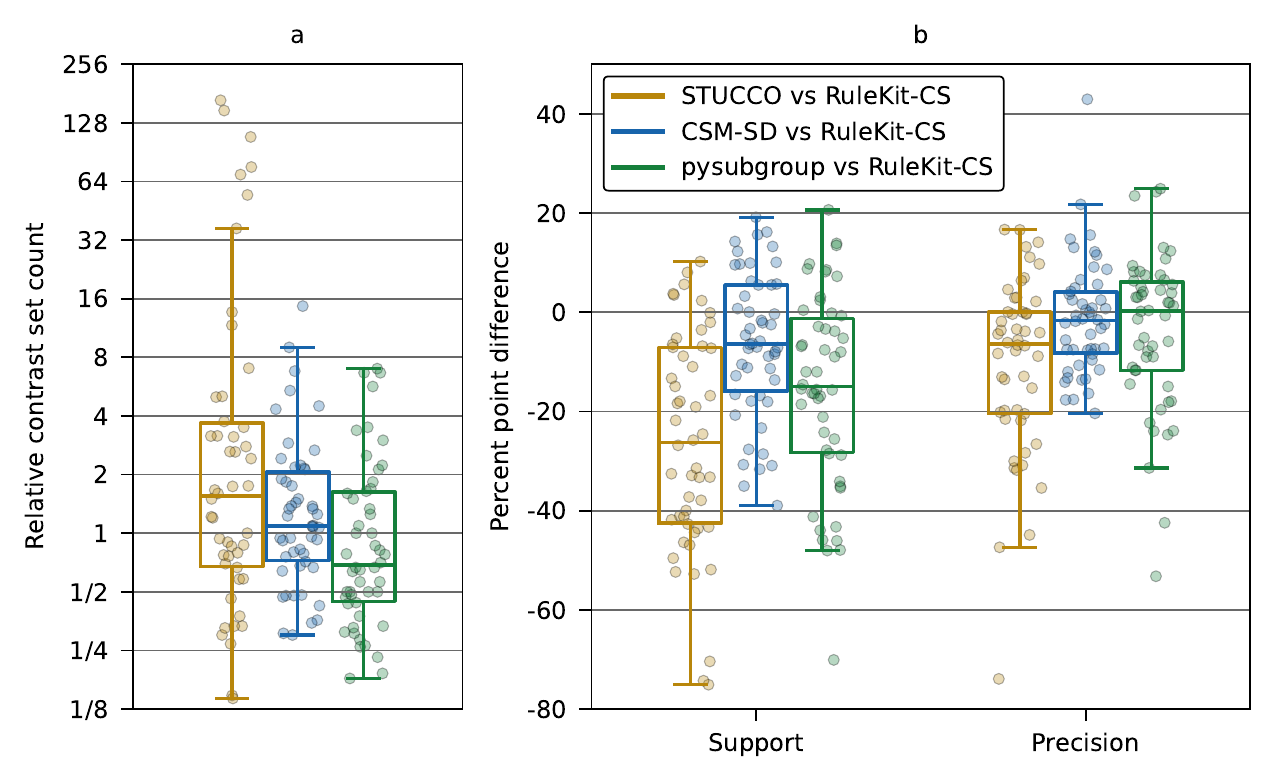}
	\caption{The comparison of the contrast set induction algorithms on the individual data sets (represented as points) at redundancy threshold 50\%. The chart presents: (a) the number of contrast sets generated by STUCCO, CSM-SD, and pysubgroup relative to \algo{}, (b) the differences of the average support and precision between STUCCO/CSM-SD/pysubgroup and \algo{}.}
	\label{fig:boxplot}
\end{figure}

The aforementioned \algo{} configuration was subject to the further analysis. To discard less original contrast set, the redundancy threshold was set to 50\%. For each data set we established: (i) the number of contrast sets generated by STUCCO, CSM-SD, and pysubgroup relative to \algo{}, (ii) the differences of the average support and precision between the competitors and our algorithm. As Figure~\ref{fig:boxplot}a shows, the large average number of contrast sets rendered by STUCCO was due to several extremely numerous cases. The distributions for CSM-SD and pysubgroup were, on the other hand, significantly less dispersed. When considering average supports (Figure~\ref{fig:boxplot}b), \algo{} was superior to all competitors, in particular to STUCCO and pysubgroup which rendered less general contrast sets in, respectively, 90\% and 80\% of the data sets. As for the precision, the advantage of our software was smaller, though still visible.     

\begin{figure}
	\centering
	\begin{subfigure}{\textwidth}
		\includegraphics{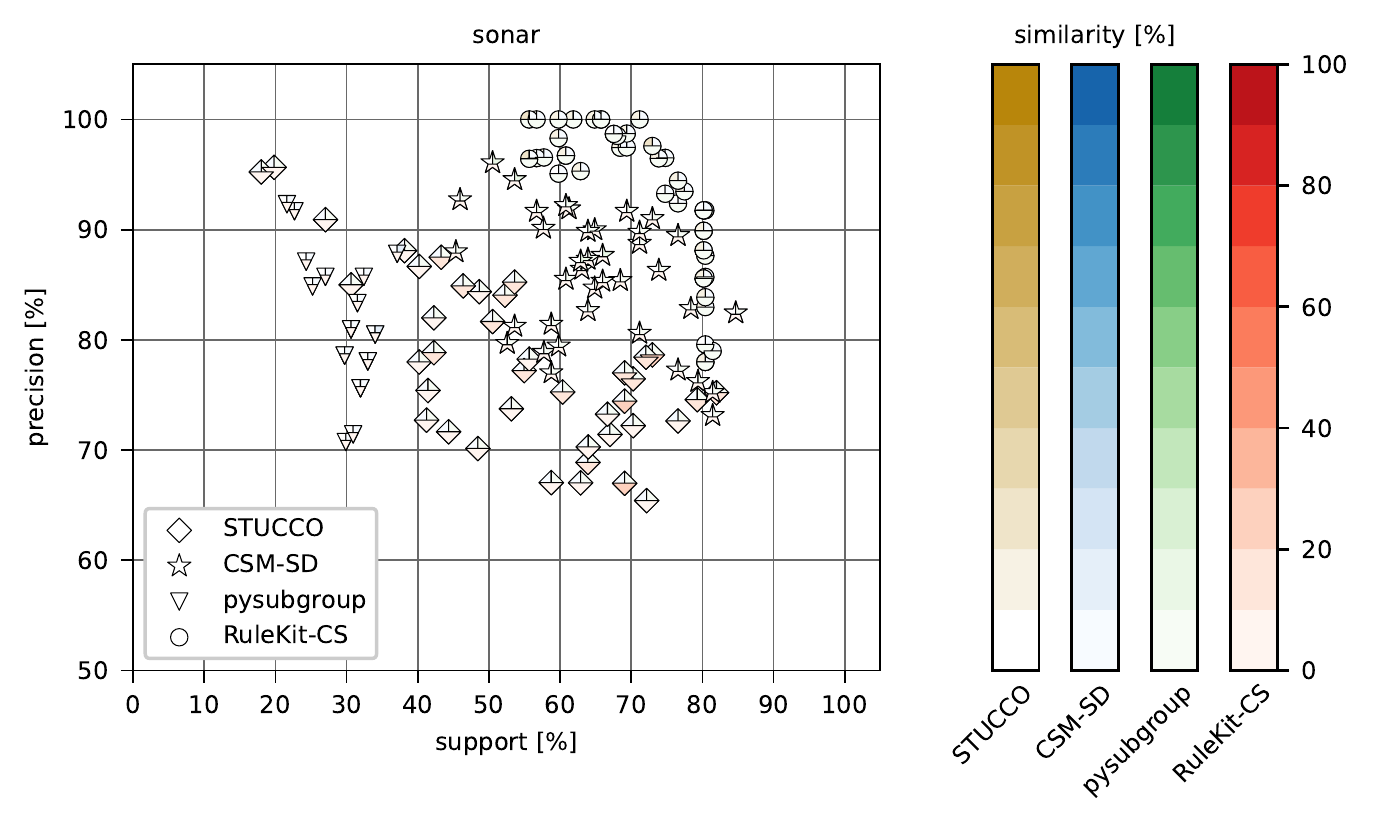}
	\end{subfigure}
	\begin{subfigure}{\textwidth}
		\includegraphics{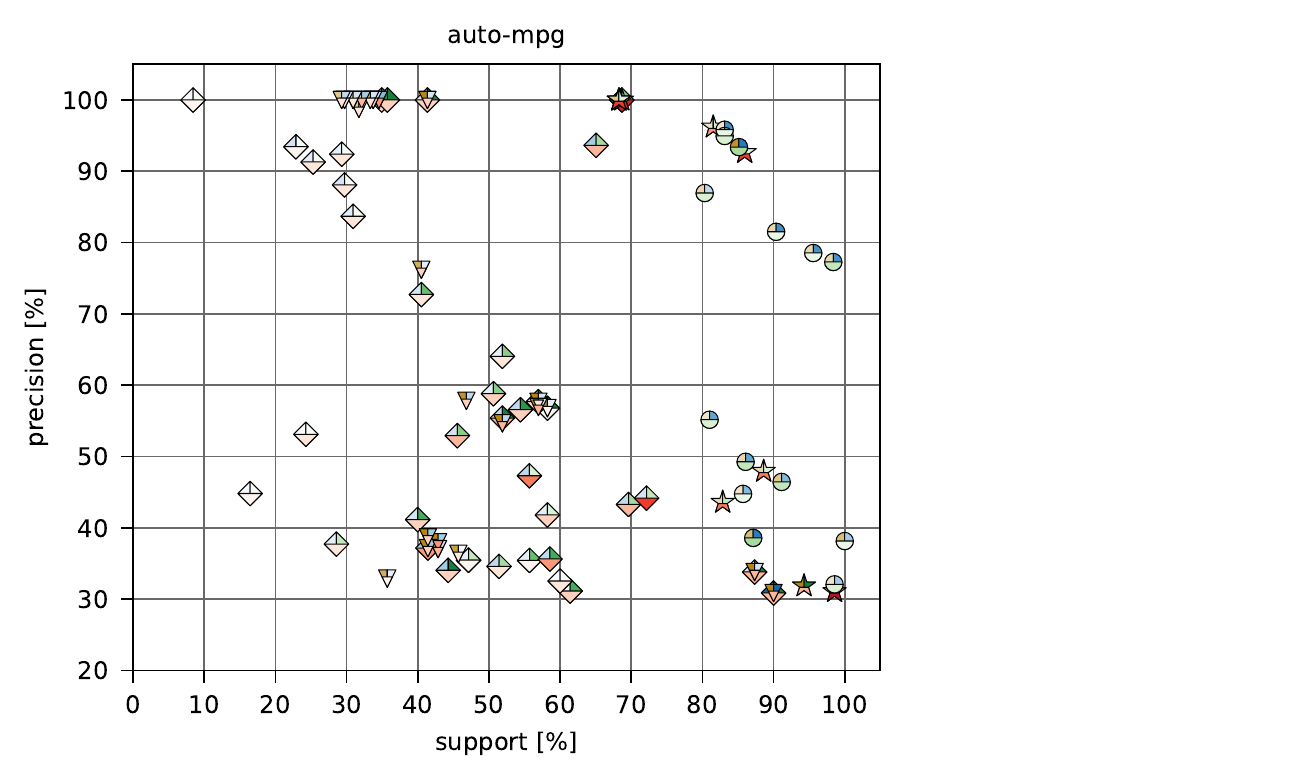}
	\end{subfigure}
	
	\caption{The visualization of individual contrast sets on support/precision plane for selected data sets at redundancy threshold 50\%. For every CS, its resemblances to the most similar contrast sets induced by the other algorithms are represented by colouring of the parts of the marker (the darker the color, the higher the similarity).}
	\label{fig:datasets}
\end{figure}

In Figure~\ref{fig:datasets} we present detailed results for selected data sets. All the contrast sets were visualized on the support/precision plane with markers representing the algorithms. Additionally, each package has the colour assigned which is used to indicate how a contrast set induced by a given algorithm resembles its most similar counterparts generated by the three other methods. For instance, a pysubgroup contrast set which does not resemble any of STUCCO and CSM-SD contrast sets and is very similar to one of the \algo{} contrast sets is indicated by a triangle which is 1/3 light brown, 1/3 light green, and 1/3 dark blue. The similarity was established analogously as the redundancy (Equation~\ref{eq:redundancy}).

As shown in Figure~\ref{fig:datasets}, the relations between CS induced by the investigated algorithms varied significantly across data sets. In the case of \textit{sonar} data set, each package rendered very unique contrast sets dissimilar to those of competitors. STUCCO, CSM-SD, and \algo{} induced similar number of CS (40, 38, and 46, respectively) with support ranging from 40 to 80\% and precision from 65 to 100\%. Interestingly, \algo{} produced the Pareto-best contrast sets --- at a given support level they were characterized by best precision and the opposite. At this background, pysubgroup was noticeably worse with only 20 contrast sets spanning 20--40\% support and 70--90\% precision range. 

The very different situation was in the case of \textit{auto-mpg} data set, where contrast sets are arranged in the several clusters with the following characteristics:
\begin{enumerate}
\setlength{\itemsep}{0pt}
\item{high support, high precision} -- small cluster dominated by \algo{}; few STUCCO and CSM-SD contrast sets which are present exhibit high similarity to those of \algo{}.
\item{high support, low precision} -- similar as (1) but with larger contrast set diversity across the algorithms,
\item{low support, high precision} -- moderately-sized cluster containing only STUCCO and pysubgroup contrast sets with mild inter-algorithm similarity,
\item{low support, low precision} -- similar as (3) but larger.  	
\end{enumerate}    

An important aspect determining usability of the presented algorithms are running times. The analysis of all 50 data sets with \algo{} on a machine equipped in Xeon E5-2670 v3 CPU ($12 \times 2.3$\,GHz cores) took 8m\,52s, making it faster then STUCCO (1h\,22m\,23s) and CSM-SD (33m\,28s) but slower then pysubgroup (37s) (note, that we decreased the value of STUCCO's \textit{maxorder} parameter for \textit{kr-vs-kp} set as the default configuration did not finish the analysis in 24 hours). Eventually, we conclude that the computational aspect does not limit the applicability of any of the investigated methods in real-life applications.

\subsection{Case study}
The analysis was performed on \textit{Statlog (Heart)} data set from UCI Machine Learning Repository. The set consists of 270 instances described by 7 numerical and 6 nominal attributes. A binary class label indicating the presence (120) or absence (150) of a heart disease was used as a group attribute. 

STUCCO and CSM-SD rendered, respectively, 21 and 54 contrast sets of which 16 and 47 had the positive support at least twice as large as the negative one ($\maxneg = 0.5$). At the same time, this condition was met for all of 20 pysubgroup and 34 \algo{} contrast sets (the latter controls this requirement during the induction). The average support and precision of the contrast sets are presented in Table~\ref{tab:summary}. An interesting observation concerns the number of examples not covered by any contrast sets. While for STUCCO, CSM-SD, and \algo{} there were no or few such cases, pysubgroup left 19 uncovered examples.      

In order to investigate the most interesting dependencies, the filtering at the redundancy threshold 50\% was performed retaining all STUCCO's and approximately two thirds of CSM-SD, pysubgroup, and \algo{} contrast sets. The average support of contrast sets observed for the presented algorithm (71.9\%) was noticeably larger than that of competitors (59.8\%, 61.9\%, and 64.2\% for STUCCO, CSM-SD, and pysubgroup, respectively). The precision of all methods was, on the other hand, similar and ranged from 75.2\% for STUCCO to 80.0\% for pysubgroup (Table~\ref{tab:summary}). Importantly, the filtering did not affect the number of uncovered examples. The increase from 3 to 6 was only observed in the number of examples covered by exactly one \algo{} contrast set. 

\begin{table}[t]
\centering
\footnotesize
\renewcommand{\tabcolsep}{0.2em}
\begin{tabular}{lp{1em}rcrcrcrcrp{1em}rcrcrcrcr}
\toprule
 && \multicolumn{9}{c}{Initial contrast sets} && \multicolumn{9}{c}{CS with redundancy $ < 50\%$} \\
		 \cline{3-11} \cline{13-21}
 			&& \# && Supp. && Prec. && $0$-cov && $1$-cov &&	\# && Supp. && Prec. && $0$-cov && $1$-cov 	\\
\midrule
STUCCO		&& 16 && 59.8 && 75.2 && 3 && 9		&& 16 && 59.8 && 75.2 && 3 && 9 \\
CSM-SD 		&& 47 && 64.6 && 77.4 && 0 && 0		&& 30 && 61.9 && 77.2 && 0 && 0	\\	
pysubgroup 	&& 20 && 61.8 && 81.3 && 19	&& 6	&& 14 && 64.2 && 80.0 && 19 && 6	\\	
\algo{} 	&& 34 && 71.7 && 78.0 && 1 && 3		&& 24 && 71.9 && 77.4 && 1 && 6 \\	
\bottomrule
\end{tabular}
\caption{\label{tab:summary}Performance of the contrast set induction algorithms on \textit{Statlog (Heart)} data set before and after redundancy filtering. Five STUCCO's and seven CSM-SD's contrast sets not fulfilling $\maxneg \leq 0.5$ condition were removed prior the analysis. Columns ``0-cov'' and ``1-cov'' indicate the numbers of examples covered by 0 and 1 contrast set, respectively.} 
\end{table}

When we investigate the presence of attributes in the resulting contrast sets, the most abundant feature was \textit{thal} which appeared in 2 STUCCO's (4\ts{th} place), 11 CSM-SD's (3\ts{rd} place), 8 pysubgroup's (1\ts{st} place), and 10 \algo{}'s (1\ts{st} place) contrast sets. The attribute indicates the presence of thalassemia (3: normal; 6: fixed defect; 7: reversible defect) and is strongly correlated with a heart disease~\citep{kremastinos2010beta}. 
Interestingly, the contrast sets based upon \textit{thal} attribute differed across algorithms (Table~\ref{tab:contrast-sets}). If we consider \textit{present} group, STUCCO induced only one such CS, which contained a single condition $\textit{thal} = 7$. This condition appeared also in all pysubgroup's contrast sets, but was absent in those generated by the other competitors. Resemblances were, however, observed in contrast sets induced by CSM-SD and \algo{}. E.g., the first contrast sets reported by these methods shared $\textit{thal} \neq 3$ and $resting\_blood\_pressure \geq 110.0$ conditions. The remaining condition was based on \textit{chest} attribute, though the value sets were different probably due to different support-precision trade offs in CSM-SD and \algo{}.

\begin{table}[h]
\centering
\footnotesize
\renewcommand{\tabcolsep}{0.2em}
\begin{tabular}{|cp{0em}|p{11cm}p{0em}|cp{0em}|c|}
	\hline
	No. && \multicolumn{1}{c}{Contrast sets for \textit{present} group} && $p$ && $n$\\
	\hline
	&& \multicolumn{1}{c}{STUCCO} && &&\\ 
	1 && $thal = 7$  && 79 && 25 \\
	\hline
	&& \multicolumn{1}{c}{CSM-SD} && &&\\ 
	1 && $thal \neq 3 \cand resting\_blood\_pressure \geq 110 \cand chest \neq 1$ && 84 && 20\\
	2 && $chest = 4 \cand oldpeak \geq 0.8 \cand thal \neq 6 \cand serum\_cholestoral \geq 164 \cand resting\_blood\_pressure \geq 108$ && 65 && \e6\\
	3 && $sex \neq 0 \cand maximum\_heart\_rate\_achieved \leq 161 \cand fasting\_blood\_sugar = 0 \cand thal \neq 6$ && 67 && 27\\
	4 && $thal \neq 3$ && 87 && 31\\ 
	5 && $maximum\_heart\_rate\_achieved \leq 156 \cand age \leq 63 \cand thal \neq 6$ && 68 && 34\\
	\hline
	&& \multicolumn{1}{c}{pysubgroup} && && \\
	1 && $thal = 7$ && 79 && 25 \\
	2 && $chest = 4 \cand thal = 7$ && 63 && \e7\\
	3 && $fasting\_blood\_sugar = 0 \cand thal = 7$ && 68 && 19\\
	4 && $sex = 1 \cand thal = 7$ && 68 && 23\\
	\hline
	&& \multicolumn{1}{c}{\algo{}} && &&\\
	1 && $chest = 4 \cand resting\_blood\_pressure \in [109, \infty) \cand thal \neq 3$ &&	67 &&	\e9 \\
	2 && $oldpeak \in [0.55, \infty) \cand thal \neq 3$ && 71 &&	13 \\
	3 && $thal \neq 3 \cand maximum\_heart\_rate\_achieved \in (-\infty, 172)$ &&	82 &&	25 \\
	4 && $serum\_cholestoral \in [145, 486.5) \cand resting\_blood\_pressure \in [109, \infty) \cand thal \neq 3$ &&	86 &&	24\\
	\hline
\end{tabular}
\caption{\label{tab:contrast-sets}The analysis of \textit{Statlog (Heart)} data set. After applying redundancy filtering at threshold 50\%, contrast sets describing \textit{present} group and containing \textit{thal} attribute were selected.} 
\end{table}

\subsection{Regression and survival contrast sets}

The ability to induce contrast sets on the regression and censored data is the unique feature of \algo{}, not provided by any other algorithm. Therefore, we investigated the differences between such contrast sets and traditional contrast sets rendered by our approach. The experiments were performed on 48 regression and 35 survival data sets from UCI repository representing wide range of problems. In the case of the regression analysis, the examples were divided into two groups: (1) those with label $L$ lower than the median of $L$ and (2) the remaining ones. For survival data the groups were defined as follows: (1) observations that were subject to an event ($L=1$) with survival time $T$ lower than the median and (2) observations with survival time greater or equal the median (both censored and uncensored). The remaining examples were removed from the sets. When inducing traditional contrast sets, label and survival time/status attributes were discarded. All the data sets in ARFF format as well as XML files with experiments configuration are available at the RuleKit repository. 

In Table~\ref{tab:reg-surv} we summarize \algo{} contrast sets induced in the mode dedicated to regression/survival data and contrast sets based on the group only. The induction was followed with the redundancy filtering at threshold 50\%. As one can see, the dedicated variant rendered 2--3 times more contrast sets as the classical mode, with the average support lower by 36 (regression) and 30 (survival) percent points. On the other hand, the precision was higher by 3 (regression) and 6 (survival) pp. The numbers of observations not covered by any contrast set were very similar for both algorithm variants. 

To get a deeper insight into differences between regression/survival and classical contrast sets, we investigated in details \textit{Bone marrow transplant: children}~\citep{sikora2019guider} survival data set. The set describes pediatric patients with hematologic diseases that were subject to the unmanipulated allogeneic unrelated donor hematopoietic stem cell transplantation. The original set consists of 187 examples characterized by 26 nominal and 9 numeric attributes (excluding survival time and status). After previously described division w.r.t. the survival time, two groups of sizes 80 and 94 were produced, while 13 examples were excluded. Attributes \textit{ANC\_recovery} and \textit{PLT\_recovery} were removed from the description due to strong correlation with a survival time.

\begin{table}[t]
	\centering
	\footnotesize
	\renewcommand{\tabcolsep}{0.18em}
	\begin{tabular}{lp{0.5em}rcrcrcrp{1em}rcrcrcr}
		\toprule
		&& \multicolumn{7}{c}{Regression/survival mode} && \multicolumn{7}{c}{Classical mode} \\
		\cline{3-9} \cline{11-17}
		Problem&& \# && Supp. && Prec. && $0$-cov && \# && Supp. && Prec. && $0$-cov \\
		\midrule
		
		Regression && $25 \pm 16$ && $38 \pm 11$ && $88 \pm 6$ && $58 \pm 111$ &&
		$9 \pm 5$ && $74 \pm 20$ && $85 \pm 7$ && $ 62 \pm 204$\\
		Survival && $22 \pm 15$ && $28 \pm \e7$ && $82 \pm 8$ && $184 \pm 252$ && $11 \pm 9$ && 
		$58 \pm 16$ && $76 \pm 10$ && $160 \pm 224$\\
		\bottomrule
	\end{tabular}
	\caption{\label{tab:reg-surv}Comparison of dedicated (regression/survival) \algo{} mode with a classical (group-only) variant on 48 regression and 35 survival data sets. The contrast sets were subject to the redundancy filtering at threshold 50\%. Column ``0-cov'' contains a number of examples not covered by any contrast set.} 
\end{table}

Running \algo{} mode dedicated to censored data induced 77 contrast sets, 54 of which fell below 50\% redundancy threshold. The average support and precision equaled to 23.6\% and 91.9\%, respectively. The classical mode produced 46 contrast sets of which 23 fulfilled the redundancy requirements. This was due to substantially higher support (61.2\%) which increased the chance of covering same examples with same attributes. The precision was, on the other hand, lower (79.8\%). Clearly, the indicators followed the general tendency presented in Table~\ref{tab:reg-surv}. 

\begin{figure}[t]
	\centering
	\includegraphics[scale=1]{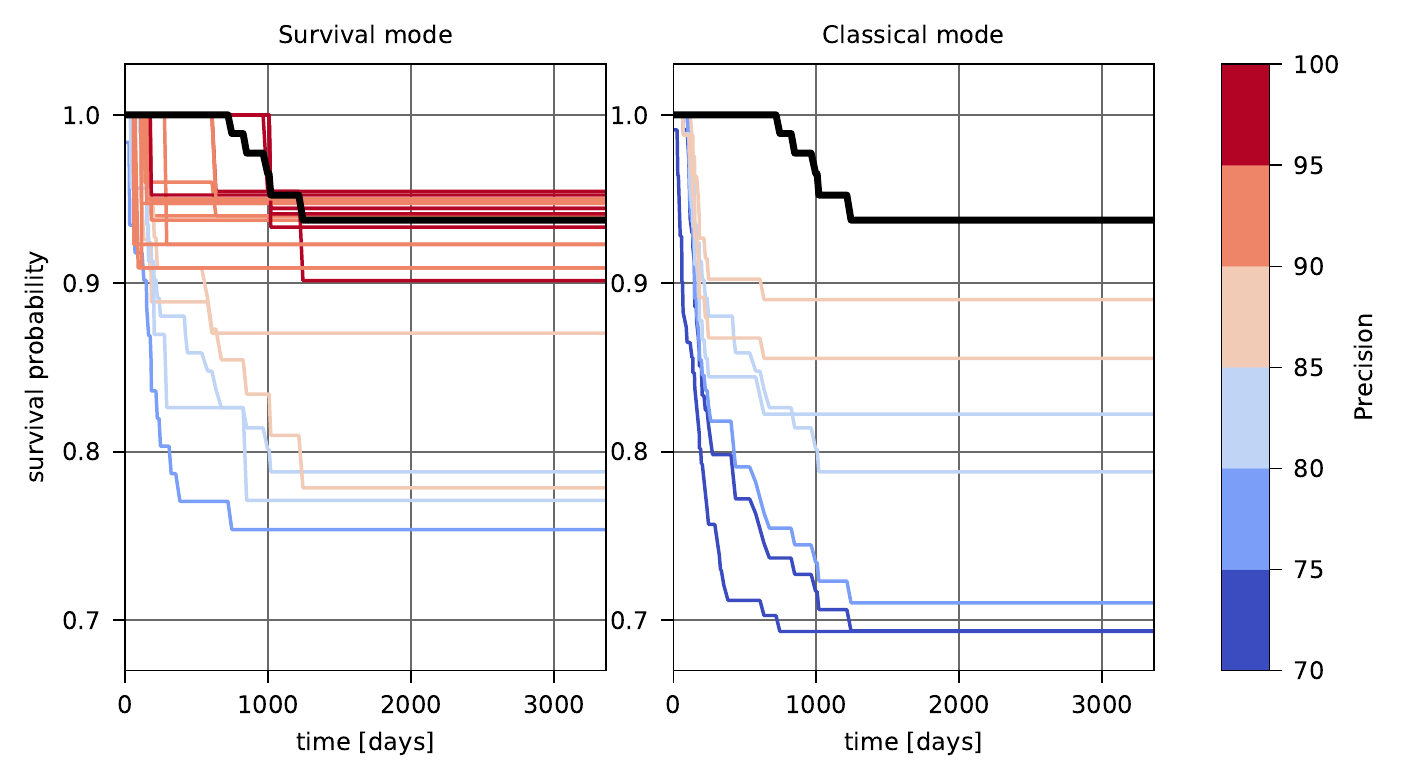}
	\caption{Comparison of survival and traditional contrast sets describing group (2) of the \textit{Bone marrow transplant: children} data set (observations with survival time greater or equal the median). The figure presents contrast sets after redundancy filtering at 50\% threshold. The survival estimate for the entire group is represented as a bold line.}
	\label{fig:survival}
\end{figure}

In Figure~\ref{fig:survival} we show survival function estimates for contrast sets (after redundancy filtering) describing group (2) -- the patients with survival time greater or equal the median (both censored and uncensored). As presented by bold line, the examples from this group have very good survival prognosis. In contrast, the examples from group (1), i.e., the patients that died in a time shorter than the median, are characterized by a pessimistic prognosis. This bimodality of survival between groups is crucial from the point of view of our algorithm. In particular, the cost of covering a negative example in the survival \algo{} mode is much larger than in the classical mode. As a result, the former induced sets of higher precision and noticeably lower support, which translated to substantially more CS in total (27 vs 7) as more contrast sets were needed to complete the covering passes. 

As presented in Figure~\ref{fig:survival}, the algorithm operated according to the expectations. In particular, the majority of the survival contrast sets (left chart) followed the survival characteristics of the entire group which was not the case in the classical mode (right chart). At the same time, the verification of \maxneg{} support criterion ensures that survival contrast sets have the basic property of contrasting groups of interest.   

The differences between \algo{} modes exhibited also in the attribute profiles. The survival contrast sets were longer then the classical ones (7.5 vs 5.2 attributes per CS) and some attributes abundant in the former, like \textit{HLA\_group\_1} or \textit{HLA\_match} (frequencies of 0.54 and 0.37, respectively) were much less frequent in the latter (0.13 and 0.13). 
This was probably due to fact, that survival prognosis represented by the survival contrast sets were more correlated with the aforementioned attributes than the main contrast group.

\section{Conclusions}
In our study we showed that sequential covering can be successfully applied for identifying contrast sets. In order to make this possible, two main challenges had to be faced:
\begin{enumerate}
\setlength{\itemsep}{0pt}
\item{ensuring the major contrast set requirement, i.e., high support in the group of interest and low support in the remaining groups,}
\item{providing the ability to identify contrast sets describing same examples but built upon different attributes.}	
\end{enumerate} 
The first aim was obtained by using correlation between predicted and target variable to drive the induction process, combined with the additional support constraints. The second point was addressed by the combination of multiple sequential covering passes and attribute penalization mechanism which ensures different induction paths between passes.
	
As presented in the experimental section, \algo{} was able to successfully analyze various data sets. When compared with CSM-SD, another approach employing sequential covering, our algorithm identified similar number of contrast sets with average support and precision superior by a small margin. When investigating particular contrast sets, the outcome strongly depended on the data set. In some cases both methods rendered similar contrast sets, in others the results differed substantially, which was due to distinct induction strategies. Therefore, if one is interested in revealing as many unknown relationships in the data as possible, a meta-analysis including \algo{} and other existing algorithms may be a wise choice.

An essential contribution of \algo{} is the ability to analyze regression and survival data by identifying  contrast sets consistent with the entire group respect to the label or survival prognosis. This type of analysis is out of reach for the existing algorithms and, as the bone marrow transplant case study shown, has the potential to discover new, interesting dependencies in the data.  

The possible improvements of \algo{} include a modification of the attribute penalization scheme. Currently, contrast set candidates are compared against global collections of already used attributes and covered examples. As a result, a candidate can be penalized even if it does not resemble any of the previous contrast sets. For instance, let $cs_1$ and $cs_2$ be two contrast sets disjoint in both, attribute and example spaces. If a candidate $cs_3$ covers same examples as $cs_1$ but using attributes from $cs_2$, it gets full penalty and no reward, which intuitively, should not be the case. A potential alternative is to verify candidates' redundancy with individual contrast sets as defined by Equation~\ref{eq:redundancy} during induction. However, tracking contrast set similarity while growing is problematic as its final form is unknown (two contrast sets may share almost entire induction path and split into disjoint sets at the very end). Moreover, the procedure would significantly increase the computational effort. Therefore, a more detailed investigation would be required to verify the advantage of this solution over currently used penalization scheme. 

RuleKit-CS was implemented as a part of RuleKit data analysis suite, which confirmed its usability in a number of problems. For instance, it has been successfully employed for detecting genetic aberrations based on the antigen expression level in B-cell precursor lymphoblastic leukemia~\cite{Kulis2022}. We believe then, that the presented algorithm could be a useful tool for identifying differences between groups with a large application potential in various areas.


\section*{Acknowledgements}
This work was supported by \L{}ukasiewicz Research Network (ROLAP\-ML, R\&D grant), The National Centre for Research and Development, Poland (POIR.01.01.01-00-0915/17), and Computer Networks and Systems Department at Silesian University of Technology within the statutory research project. We wish to thank Michael Pazzani for providing us with the implementation of STUCCO.


\bibliography{references}

\end{document}